\documentclass[aps,pra,reprint,amsmath,amssymb,longbibliography,nofootinbib,floatfix,superscriptaddress]{revtex4-1}

\usepackage{graphicx}
\usepackage{physics}
\usepackage{xcolor}
\usepackage{comment}
\usepackage{makecell}
\usepackage{hyperref}
\usepackage{bm}
\usepackage{letltxmacro}

\LetLtxMacro{\originaleqref}{\eqref}

\begin{document}

\title{Expectation Pauli-Lubanski vector and intrinsic angular momentum \\
of relativistic wavepackets}

\author{Konstantin Y. Bliokh}
\affiliation{Donostia International Physics Center (DIPC), Donostia-San Sebasti\'an 20018, Spain}
\affiliation{IKERBASQUE, Basque Foundation for Science, Bilbao 48009, Spain}

\begin{abstract}
In non-relativistic mechanics, the total (orbital) angular momentum (AM) of a spatially-distributed system can be decomposed into intrinsic and extrinsic contributions. In relativistic quantum mechanics, intrinsic AM is typically associated with spin, which can be described using the Pauli-Lubanski four-vector. 
Here, we develop a unified formalism that combines the main features of both approaches and describes the intrinsic AM of a relativistic wavepacket, including both spin and orbital contributions. 
Our approach is based on the ``expectation Pauli-Lubanski vector'' constructed from the expectation values of the wavepacket's momentum and AM. Equivalently, it defines the intrinsic AM relative to the wavepacket's energy centroid. 
In contrast to the conventional Pauli-Lubanski formalism, the zero-mass singularity does not occur for the expectation Pauli-Lubanski vector. Consequently, the intrinsic AM of a wavepacket may have an arbitrary orientation with respect to its momentum, even for massless particles. We illustrate the general theory with a number of examples of relativistic wave beams and packets carrying spin and orbital AM.  
\end{abstract}

\maketitle


\section{Introduction}

Angular momentum (AM) is a fundamental property of waves and particles, both classical and quantum. It characterizes rotational degrees of freedom and underlies an important conservation law in rotationally invariant systems.  

In classical mechanics, the AM of an extended or multi-particle system can be decomposed into {\it intrinsic} and {\it extrinsic} parts \cite{LL_mech}. The intrinsic AM quantifies rotation about the system's center of mass, whereas the extrinsic AM describes the motion of the system as a whole. 

In quantum physics, AM is separated into {\it spin} and {\it orbital} contributions \cite{LL_quantum}, with spin representing an intrinsic property of the particle. Notably, spin is a relativistic phenomenon, emerging from relativistic wave equations and field theory \cite{LL_QED, Soper_book, BB_QED}. 
At the same time, the orbital AM of transversely-confined quantum or classical wave beams or wavepackets can itself be divided into intrinsic and extrinsic parts \cite{Bliokh2015PR}, where intrinsic orbital AM can be associated with vortex-like circulating phase gradients within the packet \cite{Allen_book, Torres_book, Andrews_book, Bliokh2017PR, Karlovets2017, Ivanov2022}.

The relativistic description of AM presents additional subtleties. 
In Minkowski space-time, AM is represented by a rank-2 tensor that includes both ordinary AM and the ``boost momentum'' associated with Lorentz transformations \cite{Soper_book, LL_Field_theory}. Moreover, the spin-orbital decomposition is not Lorentz-covariant: it depends on the reference frame, and neither spin nor orbital AM are parts of four-vectors. Nevertheless, a consistent description of spin and orbital AM in relativistic quantum systems is essential for both theory and experiment. Such needs arise in optics, quantum information, and high-energy physics \cite{Bliokh2015PR, Allen_book, Torres_book, Andrews_book, Bliokh2017PR, Leader2014PR, Karlovets2017, Ivanov2022}.

A convenient theoretical tool for describing the spin states of relativistic quantum particles is the {\it Pauli-Lubanski (PL) vector} \cite{Ryder_book, Bogolubov_book}, a four-vector constructed from the operators of angular-momentum (essentially reduced to spin ${\bf S}$) and four-momentum, $p^\mu$. 
However, the conventional PL formalism does not capture the intrinsic orbital AM of wavepackets with structured phase profiles. This limitation stems from its reliance on the operator algebra, which translates directly into the corresponding eigenvalues only for {\it plane-wave} states. 
Notably, the PL formalism sharply distinguishes between massive ($p^\mu p_\mu = m^2 >0$) and massless ($p^\mu p_\mu = 0$) particles, whose AM properties differs profoundly. 

In this work, we propose a new framework that unifies three complementary perspectives: (i) the classical-mechanics decomposition of AM into intrinsic and extrinsic parts, 
(ii) the relativistic description of the AM tensor, and 
(iii) the PL formalism for relativistic quantum particles. 
The key ingredients of this approach are: 
(A) the use of the {\it energy centroid} as the reference point for the intrinsic-extrinsic decomposition, and 
(B) the use of the {\it expectation values} of momentum and AM, instead of the corresponding operators, to construct a PL-like four-vector. 
Since the PL vector involves the product of the momentum and AM operators, the product of their expectation values is {\it not} equal to the expectation value or the PL operator. 
This leads to a genuinely new quantity, which we term the {\it expectation Pauli–Lubanski (EPL) vector}.

Our approach yields several important results that distinguish it from the conventional PL formalism. 
First, the intrinsic orbital AM of structured wavepackets is naturally incorporated into the EPL vector alongside spin. 
Second, the distinction between massive and massless particles becomes far less pronounced. This is because, for any localized wavepacket, the expectation value of the four-momentum always satisfies $\langle p^\mu \rangle \langle p_\mu \rangle >0$, even when $m=0$. As a result, massless wavepackets propagate with {\it subluminal} velocities \cite{Giovannini2015, Bouchard2016O, Gouesbet2016, Bareza2016SR, Alfano2016OC, Bliokh2026} and possess a well-defined {\it rest frame}. 
Third, unlike the spin of a massless particle, which is always aligned with its momentum, the intrinsic AM (whether spin or orbital) of a wavepacket 
can be oriented arbitrarily relative to its mean momentum. 
One example is the transverse integral spin that arises in optical fields formed by superpositions of plane waves with different helicities \cite{Banzer2013, Bliokh2015PR, Bekshaev2015PRX}.
Another example is provided by optical ``spatiotemporal vortex pulses'', which carry transverse intrinsic orbital AM \cite{Bliokh2012PRA_II, Hancock2019O, Chong2020NP, Bliokh2021PRL, Hancock2021PRL, Bliokh2023PRA, Porras2023PIER, Wang2021O, Zang2022NP, Martin2025NP, Wan2023eL, Bekshaev2024APL}.

The paper is organized as follows. 
Sections~\ref{sec:mechanics} and \ref{sec:PL} review the treatment of AM in classical mechanics and relativistic quantum physics, respectively, with particular emphasis on the PL-vector formalism.
Section~\ref{sec:our_theory} introduces our general framework for intrinsic and extrinsic AM of relativistic wavepackets, together with the EPL vector. Section~\ref{sec:examples} presents illustrative examples of plane waves, beams, and wavepackets carrying spin and orbital intrinsic AM in different reference frames. Finally, Section~\ref{sec:conclusions} provides concluding remarks.

\section{Intrinsic angular momentum in non-relativistic mechanics}
\label{sec:mechanics}

We recall the main AM properties of classical nonrelativistic distributed systems \cite{LL_mech}.
Consider a set of point particles with positions ${\bf r}_i$, velocities ${\bf v}_i$, masses $m_i$,  and momenta ${\bf p}_i = m_i {\bf v}_i$. The total AM of the system is 
\begin{equation}
\label{AM_mech}
\langle {\bf J} \rangle = \Sigma_i ({\bf r}_i \times {\bf p}_i) \equiv \langle {\bf r} \times {\bf p} \rangle\,,
\end{equation} 
where we use angular brackets for integral quantities, to facilitate comparison with expectation values in quantum wave systems. 

The AM \eqref{AM_mech} generally depends on the choice of coordinate origin. Under a shift of the origin, ${\bf r}' = {\bf r} + {\bf a}$, it transforms as 
\begin{equation}
\label{AM_mech_shift}
\langle {\bf J}' \rangle = \langle {\bf J} \rangle + {\bf a} \times \langle {\bf p} \rangle\,,
\end{equation} 
where $\langle {\bf p} \rangle = \Sigma_i {\bf p}_i$ is the total momentum. If the total momentum vanishes, $\langle {\bf p} \rangle = {\bf 0}$, the AM remains invariant under origin shifts \eqref{AM_mech_shift} and may be identified as {\it intrinsic}: $\langle {\bf J}' \rangle = \langle {\bf J} \rangle \equiv {\bf J}_{\rm int}$. 

Now consider the same system in a reference frame moving with velocity ${\bf u}$ with respect to the original frame where $\langle {\bf p} \rangle = {\bf 0}$. 
In the new frame, the particle velocities become ${\bf v}'_i = {\bf v}_i - {\bf u}$, and the total AM reads
\begin{equation}
\label{AM_mech_moving}
\langle {\bf J}' \rangle =  
\langle {\bf r} \times {\bf p} \rangle - \langle m{\bf r} \times {
\bf u} \rangle = {\bf J}_{\rm int} + {\bf R} \times \langle {\bf p}' \rangle\,.
\end{equation} 
Here, ${\bf R} = \langle m {\bf r} \rangle / \langle m \rangle$
is the {\it center of mass} of the system ($\langle m \rangle$ being the total mass), and $\langle {\bf p}' \rangle = - \langle m \rangle {\bf u}$ is the total momentum of in the moving reference frame.

Thus, omitting the primes, in a generic reference frame, where $\langle {\bf p} \rangle \neq {\bf 0}$, the total AM can be decomposed into intrinsic and {extrinsic} parts, $\langle {\bf J} \rangle = {\bf J}_{\rm int} + {\bf J}_{\rm ext}$, where:
\begin{equation}
\label{AM_mech_ext_int}
{\bf J}_{\rm ext} = {\bf R} \times \langle {\bf p} \rangle\,, \quad
{\bf J}_{\rm int} = \langle {\bf J} \rangle - {\bf J}_{\rm ext}\,.
\end{equation} 
These equations may be regarded as definitions of the intrinsic and extrinsic AM \cite{Bliokh2015PR}. In particular, they show that the intrinsic AM is the AM evaluated {\it with respect to the center of mass}, i.e., when ${\bf R} = {\bf 0}$. 
Importantly, the intrinsic and extrinsic parts of the AM cannot be expressed in the ``expectation-value'' form, i.e., as sums of local single-particle quantities. Indeed, ${\bf J}_{\rm ext}$ involves the product of sums in ${\bf R}$ and $\langle {\bf p} \rangle$. Consequently, in quantum mechanics or field theory, intrinsic and extrinsic AM cannot be defined at the level of operators or local densities. 

\section{Pauli-Lubanski vector for relativistic particles}
\label{sec:PL}

We now consider relativistic quantum particles using {\it operators} underlying their evolution. We adopt the standard Minkowski-space convention with signature $(+,-,-,-)$ and units $c=1$, so that the position four-vector is $r^\mu = (t,{\bf r})$. The dynamical state of a particle is characterized by the four-momentum $p^\mu = (E,{\bf p})$ and the antisymmetric rank-2 AM tensor $J^{\mu\nu}$ \cite{LL_Field_theory, Soper_book, Ryder_book}. 
The four-momentum operator generates spacetime translations, while the AM tensor generates spacetime rotations (together they constitute 10 generators of the Poincar\'{e} group). The Minkowski-space rotations comprise ordinary spatial rotations and spatiotemporal rotations, i.e., Lorentz boosts. 
Accordingly, in a given reference frame, the AM tensor can be represented by two three-vector operators: the ordinary AM ${J}^i = \varepsilon_{ijk} J^{jk}/2$ (where $\varepsilon_{ijk}$ is the Levi-Civita symbol) and the ``boost momentum'' $K^i = J^{0i}$. 

Although the representation of the AM tensor by two three-vectors, $J^{\mu\nu} = ({\bf J}, {\bf K})$, is convenient in a chosen frame, it is not Lorentz-covariant. To provide a covariant four-vector characterization of AM, one can use the {\it Pauli-Lubanski (PL) vector} constructed from the AM tensor and four-momentum \cite{Ryder_book, Bogolubov_book}:  
\begin{equation}
\label{PL_cov}
W^\mu = \frac{1}{2} \varepsilon_{\mu\nu\rho\sigma}J^{\nu\rho} p^{\sigma} \,,
\end{equation} 
where $\varepsilon_{\mu\nu\rho\sigma}$ is the four-dimensional Levi-Civita symbol. 
The PL vector is the generator of the little group of the Poincaré group, i.e., the maximal subgroup leaving the eigenvalues of the four-momentum $p^\mu$ invariant. Note that $W^\mu p_\mu =0$ and $W^\mu W_\mu$ is Lorentz-invariant.

To examine the physical content of the PL vector, consider a specific reference frame. 
The AM tensor can be decomposed into {\it orbital} and {\it spin} parts \cite{Soper_book, LL_QED, BB_QED}: $J^{\mu\nu} = (r^\mu p^\nu - r^\nu p^\mu) + S^{\mu\nu} \equiv L^{\mu\nu} + S^{\mu\nu}$, where $S^{\mu\nu}$ is the spin tensor, and this decomposition is not Lorentz-invariant.
(In the operator description, the orbital AM and spin are often regarded as extrinsic and intrinsic, respectively, but this should not be confused with the extrinsic and intrinsic parts of the integral orbital AM discussed in Section~\ref{sec:mechanics}.)
Correspondingly, the AM vector becomes ${\bf J} = {\bf r} \times {\bf p} + {\bf S} \equiv {\bf L} + {\bf S}$, where $S^i = \varepsilon_{ijk}S^{jk}/2$ is the spin operator. In turn, the boost momentum can be written as \cite{LL_Field_theory, Soper_book, BB_QED, Ryder_book} ${\bf K} = t {\bf p} - {\bf r} E + {\bf K}_{\rm spin}$, where $K^i_{\rm spin} = S^{0i}$. The spin contribution to the boost momentum is, however, essentially auxiliary: it compensates for the non-Hermitian part of the operator ${\bf r}E$, yielding $-{\bf r}E + {\bf K}_{\rm spin} = -({\bf r} E + E {\bf r})/2$ \cite{Thaller_Dirac}. Therefore, in what follows, we assume the Hermitian symmetrization of the operator ${\bf r}E$ and omit ${\bf K}_{\rm spin}$. 
As a result, the PL vector can be written as
\begin{equation}
\label{PL_vec}
W^\mu = ({\bf p}\cdot {\bf J}, E{\bf J} - {\bf p} \times {\bf K}) 
= ({\bf p}\cdot {\bf S},E{\bf S})\,.
\end{equation} 
Thus, the PL vector encodes the {\it spin and momentum} properties of the particle. 

Importantly, the algebra of the energy-momentum and AM operators in Eq.~\eqref{PL_vec} carries over directly to the corresponding eigenvalues of {\it plane-wave} states, but it can fail to capture internal properties of structured wavepackets. Indeed, the second equality in \eqref{PL_vec} relies on the identity ${\bf L} \cdot {\bf p} = ({\bf r}\times {\bf p})\cdot {\bf p} = 0$, which holds for operators and plane waves, but fails for the expectation values of orbital AM and momentum of vortex wavepackets (which can be simultaneous eigenmodes of $L_z$ and $p_z$ with nonzero eigenvalues) \cite{Bliokh2015PR, Allen_book, Torres_book, Andrews_book, Bliokh2017PR, Karlovets2017, Ivanov2022}.

Note also that the explicit form \eqref{PL_vec} is frame dependent: spin is not itself a Lorentz-covariant quantity. Under Lorentz boosts, the AM and boost-momentum vectors mix, as do the spin and orbital AM components \cite{Smirnova2018PRA}.
A proper description of the PL vector and spin in different reference frames requires separate consideration of the massive and massless cases.

\subsection{Massive particles}
\label{sec:massive}

For a particle with nonzero mass, $m>0$, the four-momentum has the Lorentz-invariant norm
\begin{equation}
\label{p2_m}
p^\mu p_\mu = m^2> 0\,.
\end{equation} 
In the particle's rest frame, where ${\bf p}_0 = {\bf 0}$ and $E_0=m$, the spin is ${\bf S}_0$, and the PL vector \eqref{PL_vec} becomes $W^\mu_0 = (0, m {\bf S}_0)$. From this, we obtain another Lorentz-invariant quantity:
\begin{equation}
\label{W2_m}
W^\mu W_\mu = -m^2 {\bf S}_0^2 = -m^2 s(s+1)\,,
\end{equation} 
where $s$ is the spin quantum number (e.g., $s=1/2$ for a Dirac electron), and we used the spin-operator identity ${\bf S}_0^2 = s(s+1)$.

Now consider a Lorentz boost to a frame moving with velocity ${\bf u}$ relative to the rest frame. In the boosted frame, $E = \gamma m$ and ${\bf p} = -\gamma m {\bf u}$, where $\gamma = 1/\sqrt{1-u^2}$ is the Lorentz factor. In turn, the PL four-vector transforms as
\begin{equation}
\label{PL_boost}
W^\mu = m(-\gamma {\bf u}\cdot {\bf S}_0, \gamma {\bf S}_{0\parallel} + {\bf S}_{0\perp})\,,
\end{equation} 
where ${\bf S}_{0\parallel}$ and ${\bf S}_{0\perp}$ denote the components of the rest-frame spin parallel and perpendicular to ${\bf u}$, respectively. U   Using relations ${\bf S}_{0\parallel} = ({\bf S}_{0}\cdot {\bf u}){\bf u}/u^2$, ${\bf S}_{0\perp} = {\bf S}_{0} - {\bf S}_{0\parallel}$, and $E^2 - {p}^2 = m^2$, Eq.~\eqref{PL_boost} can be rewritten as
\begin{equation}
\label{PL_boost_2}
W^\mu = m\!\left(\frac{{\bf p}\cdot {\bf S}_0}{m}, {\bf S}_{0} + \frac{{\bf p}({\bf p}\cdot {\bf S}_0)}{m(E+m)} \right).
\end{equation} 

Representing the PL vector as $W^\mu = m s^\mu$, we obtain a covariant four-vector spin $s^\mu$, whose spatial part in an arbitrary reference frame is expressed via momentum, energy, and rest-frame spin as \cite{LL_QED}:
\begin{equation}
\label{spin_cov}
{\bf s} =  {\bf S}_{0} + \frac{{\bf p}({\bf p}\cdot {\bf S}_0)}{m(E+m)} \,.
\end{equation} 
This covariant spin generally differs from the ordinary non-covariant spin ${\bf S}$ (except in the rest frame, where ${\bf s} = {\bf S} = {\bf S}_0$). It is used for the relativistic vector characterization of the spin state of a moving particle. By contrast, the spin ${\bf S}$ obtained from the spin-orbital decomposition of the AM in a chosen reference frame, is obtained from Eq.~\eqref{PL_vec}:
\begin{equation}
\label{spin_vec}
{\bf S} = \frac{\bf W}{E} = \frac{m}{E} {\bf s} = \frac{m}{E} {\bf S}_{0} + \frac{{\bf p}({\bf p}\cdot {\bf S}_0)}{E(E+m)} \,.
\end{equation} 
In particular, for Dirac-electron plane waves with well-defined momentum ${\bf p}$, the expectation value of the canonical spin operator has the form of Eq.~\eqref{spin_vec} \cite{Bliokh2017PRA}.

\subsection{Massless particles}

Particles with $m=0$ have a four momentum satisfying
\begin{equation}
\label{p2_0}
p^\mu p_\mu =0\, ,
\end{equation} 
so that $E=p$. In the operator or plane-wave framework considered here, such particles possess no rest frame. If the PL vector also has zero norm,  
\begin{equation}
\label{W2_0}
W^\mu W_\mu =0\, ,
\end{equation} 
then, using the identity $W^\mu p _\mu =0$, it follows that $W^\mu = \lambda \, p^\mu$, where $\lambda$ is a Lorentz-invariant scalar.
Using Eq.~\eqref{PL_vec} in a given reference frame, we find that $\lambda = {\bf p}\cdot{\bf J}/E = {\bf p}\cdot{\bf S}/p$ is the {\it helicty}. The PL vector therefore takes the form
\begin{equation}
\label{PL_massless}
W^\mu = \left({\bf p}\cdot {\bf S}, \frac{{\bf p} ({\bf p} \cdot {\bf S})}{p}\right) .
\end{equation} 
This expression shows that only the longitudinal spin component (parallel to the momentum) is physically relevant, and the spin vector can be written as 
\begin{equation}
\label{spin_massless}
{\bf S} = \frac{\bf W}{p} = \lambda\, \frac{\bf p}{p}\, .
\end{equation} 
This spin is non-covariant: it involves the factor $1/p$ and therefore cannot be embedded into a four-vector. Unlike the massive case, Eq.~\eqref{spin_cov}, it is impossible to define covariant spin $s^\mu$, and helicity $\lambda$ is the only covariant quantity characterizing the spin state of a massless particle. This is also evident from the singular $m\to0$ limit of Eq.~\eqref{spin_cov}, whereas Eq.~\eqref{spin_vec} smoothly transforms into Eq.~\eqref{spin_massless}.

\section{Intrinsic angular momentum and expectation Pauli-Lubanski vector for relativistic wavepackets}
\label{sec:our_theory}

Let us summarize the key differences between the approaches outlined in Sections~\ref{sec:mechanics} and \ref{sec:PL}. Beyond the obvious distinction between the non-relativistic and relativistic frameworks, the classical treatment of Section~\ref{sec:mechanics} deals with  {\it distributed} systems, extended in both real and momentum space, in which purely {\it orbital} AM is decomposed into intrinsic and extrinsic parts. 
In this decomposition, the system's {\it centroid} plays a crucial role. 
By contrast, the PL-vector approach of Section~\ref{sec:PL} is formulated for {\it operators} which are straightforwardly transformed into eigenvalues of {\it plane-wave} states. Such states are spatially delocalized and therefore lack a well-defined center. Furthermore, this formalism incorporates both orbital and spin AM, yet only {\it spin} AM is intrinsic and contributes explicitly to the PL vector.

Contemporary studies of structured optical \cite{Allen_book, Torres_book, Andrews_book} or relativistic-particle \cite{Bliokh2017PR, Karlovets2017, Ivanov2022} states increasingly require an adequate description of relativistic {\it wavepackets}, i.e., sufficiently localized distributed states carrying both spin and orbital AM. 
Importantly, for wavepackets with vortex structure, the orbital AM itself can possess an intrinsic component \cite{Bliokh2015PR, Berry1998}. 
This motivates the development of a unified framework that combines the intrinsic-extrinsic AM decomposition appropriate for distributed systems with the relativistic PL-vector formalism.

\subsection{Intrinsic angular momentum}
\label{sec:intrinsic}

Consider a massive-particle wavepacket in its rest frame, which is characterized by the expectation values of energy, $\langle E_0\rangle$, and momentum, $\langle {\bf p}_0 \rangle = {\bf 0}$. Note that since the wavepacket is a superposition of multiple plane waves with different nonzero wavevectors (i.e., momenta), its mean rest-frame energy necessarily exceeds the particle mass: $\langle E_0\rangle>m$. Therefore
\begin{equation}
\label{Ep2_m}
\langle p^\mu \rangle \langle p_\mu \rangle > m^2\, .
\end{equation} 
(For wavepackets with small wavenumbers in the spectrum, one may use the approximation $\langle E_0\rangle \simeq m$ and $\langle p^\mu \rangle \langle p_\mu \rangle \simeq m^2$.)
According to the definitions of Section~\ref{sec:mechanics}, the expectation value of AM in the rest frame is purely intrinsic: $\langle {\bf J}_0 \rangle = {\bf J}_{\rm int\,0}$, and it generally includes both spin and orbital contributions: ${\bf J}_{\rm int\,0} = \langle {\bf L}_0 \rangle + \langle {\bf S}_0 \rangle$. In addition, the expectation value of the boost momentum (in an arbitrary frame) can be written as 
\begin{equation}
\label{boost_momentum}
\langle {\bf K} \rangle = t \langle {\bf p} \rangle - \langle {\bf r} E \rangle = t \langle {\bf p} \rangle - {\bf R}_E \langle E \rangle\, ,
\end{equation} 
where ${\bf R}_E = \langle {\bf r} E \rangle/\langle E \rangle$ is the {\it energy centroid} of the wavepacket \cite{LL_Field_theory, Thaller_Dirac, Smirnova2018PRA, Bliokh2012PRL}. Assuming that the energy centroid is located at the coordinate origin in the rest frame, ${\bf R}_{E\,0} = {\bf 0}$, the boost momentum vanishes there: $\langle {\bf K}_0 \rangle = {\bf 0}$.

Now perform a Lorentz boost to a frame moving with velocity ${\bf u}$. The expectation values transform in the same way as the corresponding operators, which yields \cite{Bliokh2012PRL}:
\begin{align}
\label{EV_boost}
\langle E\rangle &=\gamma \langle E_0\rangle\,, \quad
\langle {\bf p}\rangle = -\gamma \langle E_0\rangle {\bf u}\,, \nonumber \\
\langle {\bf J} \rangle &= \gamma \langle {\bf J}_{0} \rangle_\perp + \langle {\bf J}_{0} \rangle_\parallel\,, \quad
\langle {\bf K} \rangle = \gamma {\bf u} \times \langle {\bf J}_0 \rangle \,,
\end{align} 
where $\langle {\bf J}_{0} \rangle_\perp$ and $\langle {\bf J}_{0} \rangle_\parallel$ are the components of $\langle {\bf J}_0 \rangle$ perpendicular and parallel to ${\bf u}$. 
Substituting Eq.~\ref{boost_momentum} into Eqs.~\eqref{EV_boost}, we obtain the energy centroid in the boosted frame:
\begin{equation}
\label{energy_centroid}
{\bf R}_E = - {\bf u}t - \frac{{\bf u} \times \langle {\bf J}_0 \rangle}{\langle E_0 \rangle}\, .
\end{equation} 
Here, the first term describes the uniform motion of the wavepacket, whereas the second term represents the ``relativistic Hall effect'': a transverse AM-dependent shift of the wavepacket centroid induced by the Lorentz boost \cite{Bliokh2012PRL}.

The {\it extrinsic AM} of the wavepacket in the moving frame can be defined in direct analogy with the classical expression \eqref{AM_mech_ext_int}:  
\begin{equation}
\label{AM_ext}
{\bf J}_{\rm ext} = {\bf R}_E \times \langle {\bf p} \rangle = \gamma ({\bf u} \times \langle {\bf J}_0 \rangle) \times {\bf u} = \left( \gamma - \frac{1}{\gamma}\right)\!\langle {\bf J}_{0} \rangle_\perp ,
\end{equation} 
where we used the identity $u^2\gamma = \gamma - 1/\gamma$. Then, the corresponding {\it intrinsic AM} is
\begin{equation}
\label{AM_int}
{\bf J}_{\rm int} =
\langle {\bf J} \rangle - {\bf J}_{\rm ext}
= \frac{1}{\gamma}\langle {\bf J}_{0} \rangle_\perp + \langle {\bf J}_{0} \rangle_\parallel\,.
\end{equation} 
Thus, under a Lorentz boost, the transverse component of the intrinsic AM scales as $1/\gamma$, in contrast to the total AM, whose transverse component scales as $\gamma$, see Eq.~\ref{EV_boost}. 
This distinction has clear physical consequences, which will be illustrated in Section~\ref{sec:examples}.
Notably, the AM component parallel to the momentum $\langle {\bf p} \rangle$, is always intrinsic. However, the intrinsic AM ${\bf J}_{\rm int}$ generally also possesses transverse components.

In terms of the spin-orbital decomposition, $\langle {\bf J} \rangle = \langle {\bf L} \rangle + \langle {\bf S} \rangle$, the extrinsic AM \eqref{AM_ext} has a purely orbital origin. Hence, spin AM is always intrinsic, whereas orbital AM generally contains both extrinsic and intrinsic parts:
\begin{equation}
\label{int-ext-spin-orbit}
{\bf J}_{\rm ext} \equiv {\bf L}_{\rm ext}\,,\quad
\langle {\bf S} \rangle \equiv {\bf S}_{\rm int}\,, \quad
{\bf L}_{\rm int} = \langle {\bf L} \rangle - {\bf L}_{\rm ext}\,.
\end{equation} 

It is worth noticing that the above definitions of extrinsic and intrinsic AM are not unique, as they rely on the choice of the energy centroid as the reference point. Alternatively, one can use the {\it probability centroid} (or charge centroid) of the wavepacket, which dramatically affects Eqs.~\eqref{energy_centroid}--\eqref{AM_int}, as elaborated in \cite{Bliokh2012PRL, Smirnova2018PRA, Bliokh2023PRA}. However, the energy centroid is uniquely consistent with the PL formalism, as we show below.

\subsection{Expectation Pauli-Lubansky vector}
\label{sec:EPL}

We now introduce the {\it expectation Pauli-Lubansky (EPL) vector} constructed similarly to Eqs.~\eqref{PL_cov} and \eqref{PL_vec} but using the corresponding expectation values:
\begin{equation}
\label{EPL}
{\cal W}^\mu = \frac{1}{2} \varepsilon_{\mu\nu\rho\sigma}\langle J^{\nu\rho}\rangle \langle p^{\sigma}\rangle 
= (\langle {\bf p}\rangle \cdot \langle {\bf J} \rangle, \langle E \rangle \langle {\bf J} \rangle  - \langle {\bf p}\rangle  \times \langle {\bf K}\rangle )\,. 
\end{equation} 
We emphasize that ${\cal W}^\mu \neq \langle {W}^\mu\rangle$. 
In particular, the EPL vector contains both {spin and orbital} AM contributions. Since the ordinary PL vector characterizes relativistic spin, it is natural to interpret the EPL vector as the relativistic descriptor of the intrinsic AM of a wavepacket. 

For a massive wavepacket considered above, the EPL vector in the rest frame reads
\begin{equation}
\label{EPL_rest}
{\cal W}^\mu_0  
= (0, \langle E_0 \rangle \langle {\bf J}_0 \rangle)\,,
\end{equation} 
which is analogous to $W^\mu_0 = (0, m {\bf S_0})$ in Section~\ref{sec:massive}. Hence, the Lorentz-invariant norm of the EPL vector is 
\begin{equation}
\label{EW^2_m}
{\cal W}^\mu {\cal W}_\mu = - \langle E_0 \rangle^2 \langle {\bf J}_0 \rangle^2\,.
\end{equation} 
Applying a Lorentz boost to Eq.~\eqref{EPL_rest} yields the EPL vector in the moving frame:
\begin{equation}
\label{EPL_moving}
{\cal W}^\mu  
= \langle E_0 \rangle (-\gamma {\bf u}\cdot \langle {\bf J}_0 \rangle, \langle {\bf J}_0 \rangle_{\perp} + \gamma \langle {\bf J}_0 \rangle_{\parallel})\,.
\end{equation} 
The same expression follows by substituting the transformed expectation values \eqref{EV_boost} into Eq.~\eqref{EPL}.

Equation~\eqref{EPL_moving} has the same structure as the boosted PL vector in Eq.~\eqref{PL_boost}. Therefore, similarly to Section~\ref{sec:massive}, we can introduce the covariant four-vector of intrinsic AM: $j^\mu_{\rm int} = {\cal W}^\mu/\langle E_0 \rangle$. Its spatial part in an arbitrary reference frame in takes the form similar to Eq.~\eqref{spin_cov}: 
\begin{align}
\label{AM_int_cov}
{\bf j}_{\rm int} & = \frac{\bm{\mathcal W}}{\langle E_0\rangle} 
=  \langle {\bf J}_0 \rangle_{\perp} + {\gamma}\langle {\bf J}_0 \rangle_{\parallel} \nonumber \\
& =  \langle {\bf J}_{0} \rangle + \frac{\langle {\bf p} \rangle (\langle {\bf p} \rangle \cdot \langle {\bf J}_0 \rangle )}{\langle E_0 \rangle(\langle E \rangle + \langle E_0 \rangle)} \,.
\end{align} 
Furthermore, defining the non-covariant intrinsic AM similar to Eq.~\eqref{spin_vec} yields:
\begin{align}
\label{AM_int_vec}
{\bf J}_{\rm int} & = \frac{\bm{\mathcal W}}{\langle E\rangle} 
=  \frac{1}{\gamma}\langle {\bf J}_0 \rangle_{\perp} + \langle {\bf J}_0 \rangle_{\parallel} \nonumber \\
&=\frac{\langle E_0 \rangle}{\langle E\rangle} \langle {\bf J}_{0} \rangle + \frac{\langle {\bf p} \rangle (\langle {\bf p} \rangle \cdot \langle {\bf J}_0 \rangle )}{\langle E\rangle(\langle E \rangle +\langle E_0 \rangle)} \,.
\end{align} 

Remarkably, Eq.~\eqref{AM_int_vec} coincides exactly with Eq.~\eqref{AM_int}, which was derived from classical-mechanics-like arguments based on the energy centroid ${\bf R}_E$. Thus, the EPL-vector formalism naturally unifies the classical intrinsic-extrinsic AM decomposition and the PL description of relativistic spin. This constitutes the central result of the present work. 
In Section~\ref{sec:examples}, we illustrate the physical content of this formalism by several explicit examples.

\subsection{Massless case}

Remarkably, our formalism applies equally well to wavepackets of massless particles. Unlike plane waves (which are not physically realizable states), localized massless wavepackets can never propagate at the speed of light. 
Their spatial confinement is provided by the interference of multiple plane waves in the wavepacket spectrum, which have slightly different propagation directions. This results in {\it subluminal} propagation of the wavepacket -- a phenomenon which is well understood and has been measured for photons \cite{Giovannini2015, Bouchard2016O, Gouesbet2016, Bareza2016SR, Alfano2016OC, Bliokh2026}. This means that the expectation value of the four-momentum has a non-zero timelike norm, Eq.~\eqref{Ep2_m}:    
\begin{equation}
\label{Ep2_0}
\langle p^\mu  \rangle \langle p_\mu  \rangle >0 \,.
\end{equation} 
One can say that a wavepacket possesses an ``{\it effective mass}'' $\langle p^\mu  \rangle \langle p_\mu  \rangle = m^2_{\rm eff}$ \cite{Alfano2016OC, Karlovets2018PRA, Silenko2019PRA}, which depends on wavepacket structure (in particular, on its spatial size). 

Moreover, any subluminal wavepacket admits a well-defined {\it rest frame}, where the expectation value of the momentum vanishes: $\langle {\bf p}_0  \rangle = {\bf 0}$. In this frame, the wavepacket becomes a superposition of plane waves propagating in opposite directions, i.e., a standing wave (see the examples in Sections~\ref{sec:transverse_spin} and \ref{sec:Bessel_long}). The effective mass is then equal to the mean rest-frame energy: $m_{\rm eff} = \langle E_0  \rangle$.  

Thus, the massless-wavepacket case becomes formally equivalent to the massive case, and all results derived in Sections~\ref{sec:intrinsic} and \ref{sec:EPL} remain valid without modification. 
This is in sharp contrast to the PL vector for plane waves, which is singular at $m=0$. 
Consequently, unlike the spin of a massless plane wave, Eq.~\eqref{spin_massless}, which is always aligned with the momentum, the intrinsic AM of a massless wavepacket, ${\bf J}_{\rm int}$, can have an arbitrary orientation with respect to its momentum $\langle {\bf p} \rangle$. This has already been demonstrated in optics. 
First, the integral spin AM $\langle {\bf S} \rangle$ of optical fields composed of plane waves with different helicities generally acquires a component transverse to the mean momentum \cite{Banzer2013, Bliokh2015PR, Bekshaev2015PRX} (see Section~\ref{sec:transverse_spin}).
Second, spatiotemporal optical vortex wavepackets \cite{Bliokh2012PRA_II, Hancock2019O, Chong2020NP, Bliokh2021PRL, Hancock2021PRL, Bliokh2023PRA, Porras2023PIER, Wang2021O, Zang2022NP, Martin2025NP, Wan2023eL, Bekshaev2024APL} can carry intrinsic orbital AM, ${\bf L}_{\rm int}$, that is transverse or tilted with respect to their propagation direction (see Section~\ref{sec:KG}). 

\section{Examples}
\label{sec:examples}

\subsection{Plane electromagnetic wave}
\label{sec:plane}

\begin{figure}[t!]
\centering
\includegraphics[width=\linewidth]{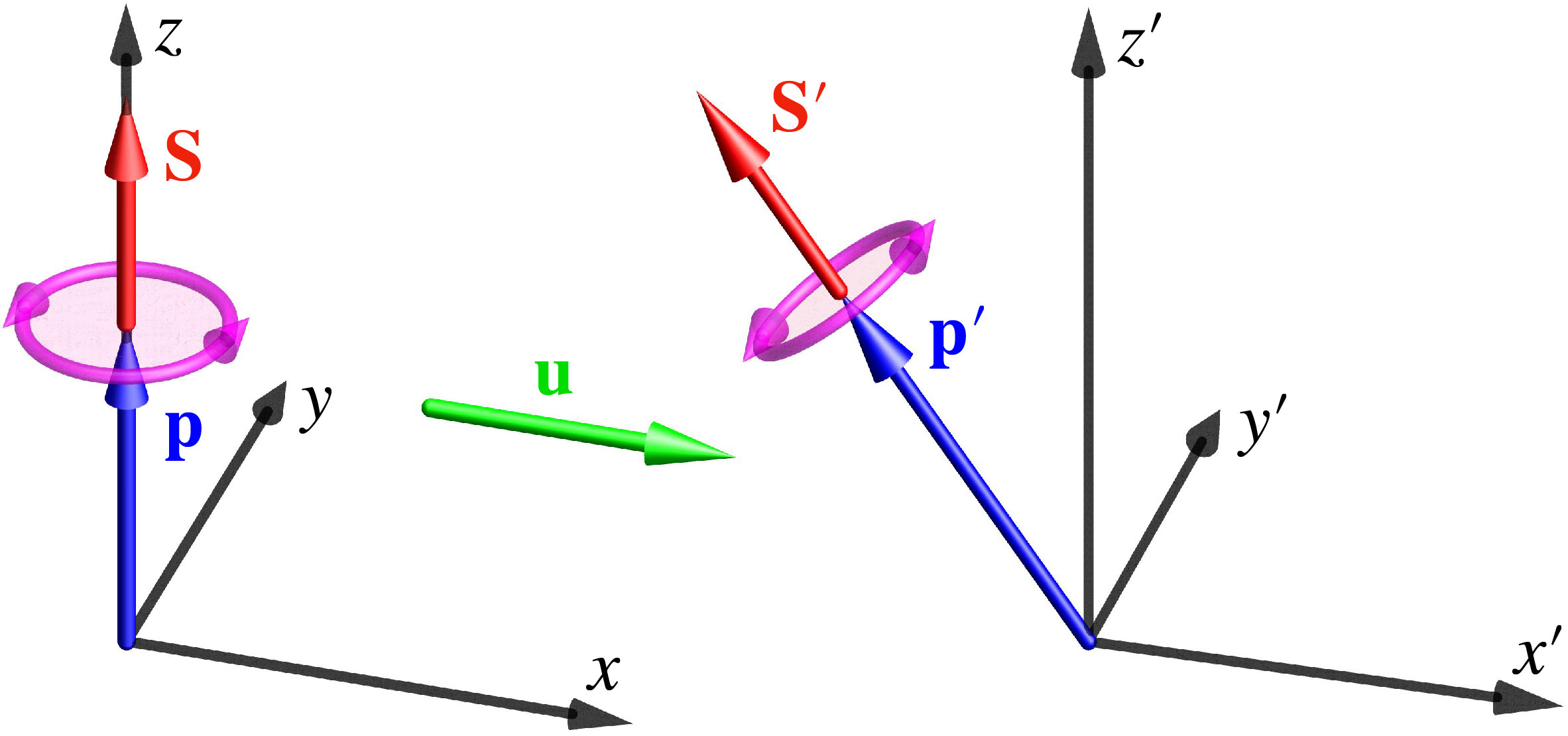}
\caption{Transverse Lorentz boost with velocity ${\bf u} =0.6 \hat{\bf x}$ applied to a $z$-propagating circularly-polarized plane electromagnetic wave (shown here for helicity $\lambda =1$). The momentum, spin, and polarization are shown by the blue, red, and magenta arrows, respectively. The spin remains aligned with the momentum, see Eqs.~\eqref{spin_massless} and \eqref{pw_spin}.}
\label{Fig_1}
\end{figure}

We begin with a simple example previously mentioned in \cite{Smirnova2018PRA}. Let a circularly-polarized plane electromagnetic wave propagate in free space along the $z$-axis, see Fig.~\ref{Fig_1}. The AM state of this wave is described by spin ${\bf S} = \lambda \hat{\bf z}$, where $\lambda = \pm 1$ is the helicity, which corresponds to the right-hand and left-hand circular polarizations (we use the $\hbar=1$ units and normalize dynamical quantities per photon per unit volume). The energy and momentum of this plane wave are quantified by its frequency $\omega$ and wavenumber $k$: $E = \omega = k$ and ${\bf p} = k \hat{\bf z}$.

Now consider a Lorentz boost to a frame moving with velocity ${\bf u} = u \hat{\bf x}$. In the boosted frame, the wave retains its cricular polarization (i.e., helicity), but its energy and momentum become $E' = \gamma k$ and ${\bf p}' =  k \hat{\bf z} - \gamma u k \hat{\bf x}$. Accordingly, the spin transforms to
\begin{equation}
\label{pw_spin}
{\bf S}' = \lambda \frac{{\bf p}'}{p'}= 
\lambda \!\left( \frac{1}{\gamma} \hat{\bf z} - u\hat{\bf x}\right) ,
\end{equation} 
which agrees with the massless-particle relation \eqref{spin_massless}. Notably, the transverse (with respect to ${\bf u}$) $z$-component of the spin scales as $1/\gamma$. This differs from the transformation of the total AM with the factor of $\gamma$ in Eqs.~\eqref{EV_boost}, but agrees with the transformation of the intrinsic AM in Eq.~\eqref{AM_int}. This apparent discrepancy is resolved once one considers a localized wavepacket and expectation values instead of a single plane wave \cite{Smirnova2018PRA}. 

\subsection{Transverse spin in two-wave interference}
\label{sec:transverse_spin}

\begin{figure}[t!]
\centering
\includegraphics[width=\linewidth]{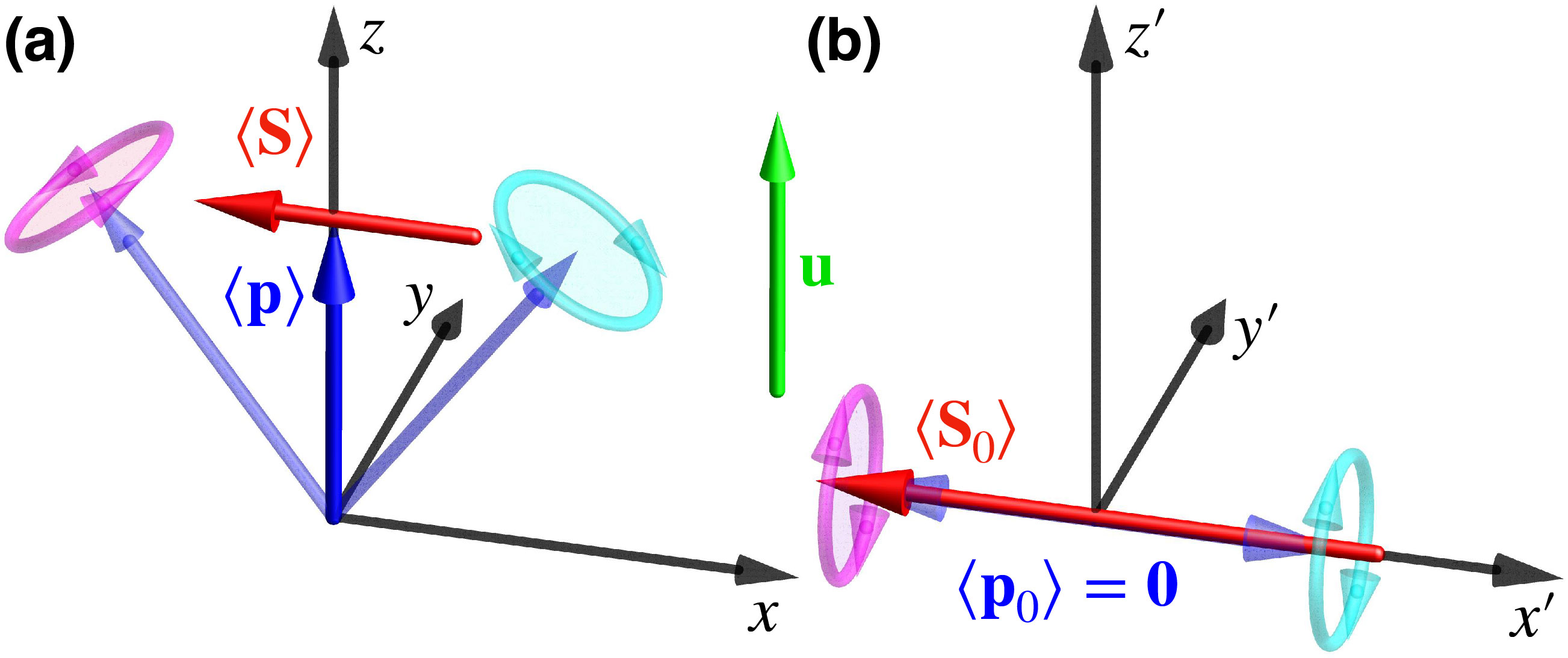}
\caption{(a) Interference of two plane electromagnetic waves with different wavevectors and opposite helicities (shown in magenta and cyan) yields a field with the mean spin (intrinsic AM) $\langle {\bf S}\rangle$ orthogonal to the mean momentum $\langle {\bf p}\rangle$ \cite{Banzer2013, Bliokh2015PR, Bekshaev2015PRX}. (b) A Lorentz boost with velocity ${\bf u} = \langle {\bf p}\rangle/ \langle E \rangle$ transforms this field to its rest frame, where $\langle {\bf p}_0\rangle = {\bf 0}$ and the field becomes a standing wave with the mean spin $\langle {\bf S}_0\rangle = \gamma \langle {\bf S}\rangle$.}
\label{Fig_two_waves}
\end{figure}

The interference of just two electromagnetic plane waves provides the simplest example of the transverse intrinsic AM. Let the two waves have wavevectors ${\bf k}_{1,2} = k_z \hat{\bf z} \pm k_x \hat{\bf x}$ and opposite helicities $\lambda_{1,2} = \mp 1$. Although the resulting field is not spatially localized, its properly normalized mean energy, momentum and spin are readily calculated \cite{Banzer2013, Bliokh2015PR, Bekshaev2015PRX} (see Fig.~\ref{Fig_two_waves}): 
\begin{align}
\label{two-wave}
\langle E \rangle & = k\,, \quad\langle {\bf p} \rangle  = \frac{{\bf k}_1+ {\bf k}_2}{2}= k_z \hat{\bf z} \,, \nonumber \\
\langle {\bf S} \rangle & = \frac{\lambda_1{\bf k}_1+ \lambda_2{\bf k}_2}{2k} = - \frac{k_x}{k} \hat{\bf x}\,.
\end{align} 
Thus, the mean spin is purely transverse to the mean momentum. This sharply contrasts with the plane-wave result for a massless particle, Eqs.~\eqref{spin_massless} and \eqref{pw_spin}.

Importantly, this two-wave system admits a rest frame, see Fig.~\ref{Fig_two_waves}. A Lorentz boost with velocity ${\bf u} = (k_z/k)\hat{\bf z}$ and $\gamma = k/k_x$ transforms the wave parameters into $\omega' = k_x$ and ${\bf k}'_{1,2} = \pm k_x \hat{\bf x}$, while preserving their helicities: $\lambda_{1,2}'=\lambda_{1,2}$. In this frame, the field becomes a standing wave with mean characteristics
\begin{align}
\label{two-wave-rest}
\langle E_0 \rangle = k_x\,, \quad
\langle {\bf p}_0 \rangle  = {\bf 0} \,, \quad
\langle {\bf S}_0 \rangle  = - \hat{\bf x} = \gamma \langle {\bf S} \rangle\,.
\end{align} 
The transformation of the mean spin between Eqs.~\eqref{two-wave-rest} and \eqref{two-wave} exactly matches the Lorentz transformation of intrinsic AM, Eq.~\eqref{AM_int}. Moreover, although the two-wave field is not sufficiently localized for a rigorous definition of the EPL vector, Eqs.~\eqref{EPL}--\eqref{EPL_moving}, (the expectation values of the boost momentum and energy centroid are ill-defined), the above results are fully consistent with the EPL vector 
\begin{align}
\label{two-wave-EPL}
{\mathcal{W}}^\mu = {\mathcal{W}}_0^\mu = (0, -k_x \hat{\bf x})\,, ~~
{\mathcal{W}}^\mu {\mathcal{W}}_\mu = \langle E_0 \rangle^2 \langle {\bf S}_0 \rangle^2= k_x^2 \,,
\end{align} 
and relation \eqref{AM_int_vec} $\langle {\bf S} \rangle = \bm{\mathcal W}/ \langle E \rangle$. 

\subsection{Bessel beam and its rest frame}
\label{sec:Bessel_long}

Next, we consider a transversely-localized Bessel beam carrying both spin and orbital AM. Bessel-beam states can be constructed for both massless (e.g., electromagnetic) \cite{McGloin2005CP, Bliokh2010PRA} and massive (e.g., electron) \cite{Bliokh2011PRL, Bliokh2017PR} waves. Although such beams are not properly confined wavepackets (they are delocalized along the propagation direction and are not square-integrable in the transverse plane), they 
provide a transparent illustration of the main ideas developed here. 

\begin{figure}[t!]
\centering
\includegraphics[width=\linewidth]{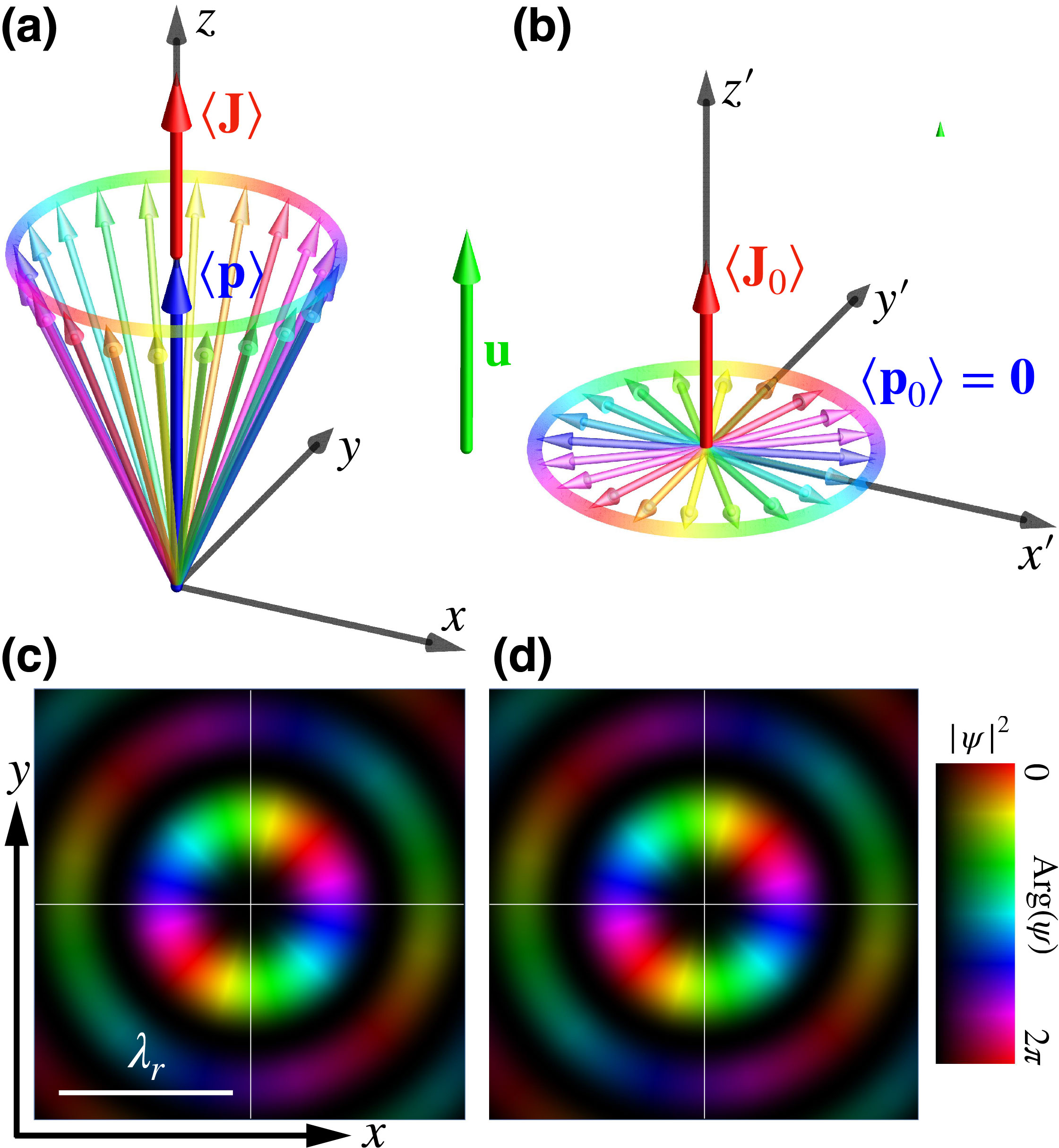}
\caption{(a) Plane-wave spectrum of a Bessel beam with $k_r =0.4 k$, arbitrary $m$, and azimuthal phase winding corresponding to $\ell=2$ (encoded by hue colors). The expectation values of the momentum and intrinsic AM are shown by the large blue and red arrows, respectively. 
(b) Lorentz boost to the rest frame (${\bf u} = \hat{\bf z} k_z/\omega$), where the mean momentum vanishes, while the intrinsic AM remains unchanged, see Eqs.~\eqref{EV_Bessel} and \eqref{EV_Bessel_rest}. 
(c,d) Corresponding transverse spatial distributions of the scalar wavefunction $\psi({\bf r})$ (obtained by Fourier transform of the plane-wave spectra). Brightness and hue colors represent intensity and phase, respectively. 
The distributions are identical in the two frames. The scale bar corresponds to $\lambda_r = 2\pi/k_r$.}
\label{Fig_2}
\end{figure}

A $z$-propagating Bessel beam is a superposition of plane waves whose wavevectors are uniformly distributed along a circle around the $k_z$ axis: ${\bf k} \equiv (k_x,k_y,k_z) = (k_z, k_r \cos\phi, k_r \sin\phi)$, where $\phi \in [0,2\pi)$ is the azimuthal coordinate in the ${\bf k}$ space, and $k_z^2 +k_r^2 =k^2$, see Fig.~\ref{Fig_2}(a). It is also convenient to use the polar angle $\theta$: $k_z = k \cos\theta$, $k_r = k \sin\theta$; for fixed $k$, it determines the transverse size of the beam.

The plane-wave components may also carry an azimuthal phase factor $e^{i\ell\phi}$, where $\ell \in \mathbb{Z}$ is an integer number. This endows the beam with a phase vortex and corresponding orbital AM determined by $\ell$ \cite{Bliokh2015PR, Allen_book, Torres_book, Andrews_book, Bliokh2017PR}. Including spin polarization characterized by the quantum number $\sigma$, we consider a beam that is an eigenstate of the $z$-component of the total AM, with eigenvalue $(\ell + \sigma)$. For photons, $\sigma = \lambda = \pm 1$ is the helicity of each plane-wave component \cite{Bliokh2010PRA}, while for electrons $\sigma = s_z = \pm 1/2$ can be associated with the $z$-component of the rest-frame spin for each plane wave \cite{Bliokh2011PRL}.

The expectation values of energy, momentum, and total AM read (in suitable units):
\begin{equation}
\label{EV_Bessel}
\langle E \rangle = \omega\,, \quad  \langle {\bf p} \rangle = k_z \hat{\bf z}\,, \quad
\langle {\bf J} \rangle = (\ell+\sigma)\hat{\bf z}\,.
\end{equation} 
Here $\omega = \sqrt{m^2+ k^2}$ is the frequency which can describe both massive and massless cases.
For this Bessel beam, the transverse components of the mean boost momentum and  of the energy centroid vanish: $\langle {\bf K}\rangle \times \langle {\bf p}\rangle = {\bf R}_E \times  \langle {\bf p}\rangle = {\bf 0}$.
The longitudinal AM in Eq.~\eqref{EV_Bessel} is purely intrinsic, $\langle {\bf J} \rangle = {\bf J}_{\rm int}$, and its value is independent of the choice of coordinate origin \cite{Bliokh2015PR, Berry1998}. 
Its value is also independent of the polar angle $\theta$, although the spin-orbital decomposition $\langle {\bf J} \rangle = \langle {\bf L} \rangle + \langle {\bf S} \rangle$ does depend on $\theta$ \cite{Bliokh2010PRA, Bliokh2011PRL}. 

The Bessel beam possesses an effective mass: 
\begin{equation}
\label{m_eff}
\langle p^\mu  \rangle \langle p_\mu  \rangle = \omega^2 - k_z^2 = m^2 +k^2 \sin^2\!\theta= m^2_{\rm eff}\,, 
\end{equation} 
which remains nonzero even for $m=0$, in agreement with Eqs.~\eqref{Ep2_m} and \eqref{Ep2_0}. Consequently, one can perform a Lorentz boost with velocity ${\bf u} = \hat{\bf z}\, k_z/\omega$ to the rest frame of the beam. In this frame:
\begin{equation}
\label{EV_Bessel_rest}
\langle E_0 \rangle = m_{\rm eff}\,, ~~  \langle {\bf p}_0 \rangle = {\bf 0}\,, ~~
\langle {\bf J}_0 \rangle = (\ell+\sigma)\hat{\bf z}\,.  
\end{equation} 
The plane-wave spectrum becomes a circle in the $(k_x,k_y)$ plane, see Fig.~\ref{Fig_2}(b). The invariance of the longitudinal intrinsic AM, $\langle {\bf J}_0 \rangle =\langle {\bf J} \rangle$, under a longitudinal Lorentz boost is consistent with Eqs.~\eqref{AM_int} and \eqref{AM_int_vec}. It differs, however, from the behavior of the covariant intrinsic AM ${\bf j}_{\rm int}$, Eq.~\eqref{AM_int_cov}, whose longitudinal component scales by a factor of $\gamma$.

The EPL vector \eqref{EPL}--\eqref{EPL_moving} for the Bessel beam in the initial (moving) and rest frames reads:
\begin{align}
\label{EPL_Bessel}
{\cal W}^\mu & = 
(\ell+\sigma)\left(k_z , \omega \hat{\bf z} \right), \nonumber \\
{\cal W}^\mu_0 & = 
(\ell+\sigma)\left(0, m_{\rm eff}\,\hat{\bf z} \right).
\end{align} 
Unlike the conventional PL vector, this EPL vector includes both the spin and orbital contributions to the intrinsic AM.  It agrees with the general relation \eqref{AM_int_vec}, ${\bf J}_{\rm int} = \bm{\mathcal W}/ \langle E \rangle$, and its Lorentz-invariant norm \eqref{EW^2_m} is: 
\begin{equation}
\label{EW2_Bessel}
{\cal W}^\mu{\cal W}_\mu = 
-m_{\rm eff}^2(\ell+\sigma)^2\,.  
\end{equation} 
%

\subsection{Transverse boosts of a paraxial Bessel beam}
\label{sec:Bessel_trans}

We now examine another instructive example, previously considered in \cite{Smirnova2018PRA}. In that work, the intrinsic and extrinsic AM were defined with respect to the {probability} centroid. Here we revisit the same system using our present formalism based on the {energy} centroid.

Consider the Bessel beam \eqref{EV_Bessel} in the paraxial approximation $\theta\ll 1$, $k_z\simeq k$, $m_{\rm eff} \simeq m$. In this regime, the spin-orbit coupling is negligible, and the spin and orbital parts of the intrinsic AM are: $\langle {\bf L} \rangle \simeq \ell \hat{\bf z}$ and $\langle {\bf S} \rangle \simeq \sigma  \hat{\bf z}$ \cite{Bliokh2015PR, Allen_book, Bliokh2017PR}.

\begin{figure}[t!]
\centering
\includegraphics[width=\linewidth]{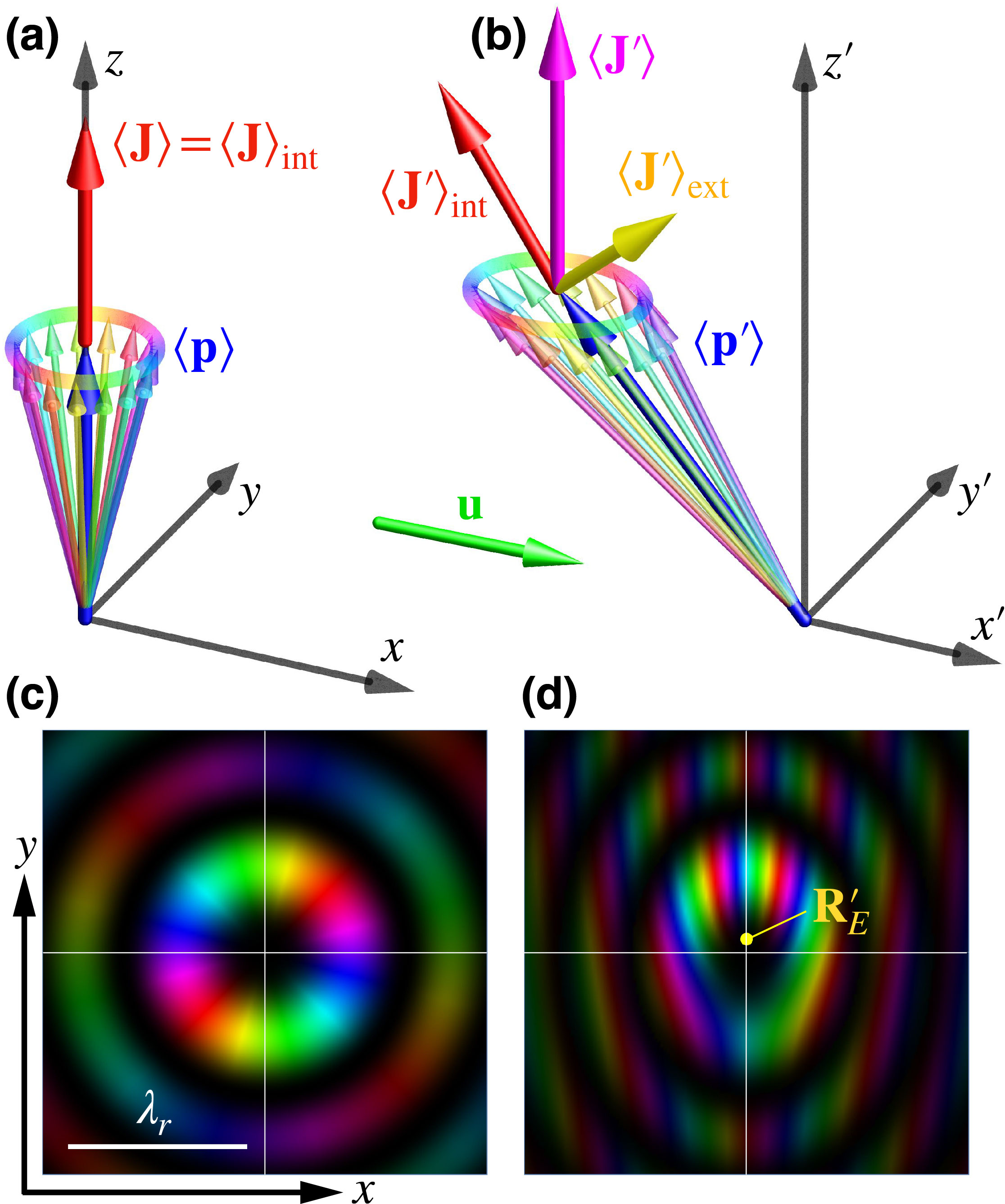}
\caption{(a) Plane-wave spectrum of a paraxial Bessel beam with $m=0.5 \omega$, $k_r = 0.2 k$, and $\ell=2$. 
(b) Transverse Lorentz boost with velocity ${\bf u} = 0.6 \hat{\bf x}$. 
The total, intrinsic, and extrinsic AM, given by Eqs.~\eqref{EV_Bessel_moving}--\eqref{AM_int_Bessel}, are shown by the magenta, red, and yellow arrows, respectively. The intrinsic AM is aligned with the momentum only in the massless case, $m=0$. 
(c,d) Transverse spatial distributions of the corresponding wave fields in the original and boosted frames. Brightness and hue colors represent the energy density and phase, respectively (for simplicity, the energy density was determined for the scalar Klein-Gordon field, as in \cite{Bliokh2012PRL, Bliokh2012PRA_II}). The transverse shift of the energy centroid, responsible for the extrinsic AM in the boosted frame, is shown in (d) by the yellow dot.}
\label{Fig_3}
\end{figure}
    
Let us perform a Lorentz boost to a frame moving with velocity ${\bf u} = u \hat{\bf x}$, $u \gg \theta$. In this frame, the beam becomes a spatiotemporal vortex beam \cite{Bliokh2012PRA_II, Smirnova2018PRA}, as shown in Fig.~\ref{Fig_3}. The expectation values and the energy centroid become \cite{Smirnova2018PRA}:
\begin{align}
\label{EV_Bessel_moving}
\langle E' \rangle & = \gamma \omega \,, \quad  
\langle {\bf p}' \rangle \simeq k \hat{\bf z}- \gamma u \omega\hat{\bf x}\,, \quad
\langle {\bf J}' \rangle = \gamma (\ell+\sigma) \hat{\bf z},
\nonumber \\
\langle {\bf L}' \rangle & \simeq \gamma\!\left( \ell + \sigma u^2 \right)\!\hat{\bf z} + \sigma\frac{k}{\omega} u \hat{\bf x}, ~~
\langle {\bf S}' \rangle  \simeq \sigma\! \left( \frac{1}{\gamma} \hat{\bf z} - u \frac{k}{\omega}\hat{\bf x}\right)\!, \nonumber \\
\langle {\bf K}' \rangle & \simeq - \gamma u (\ell + \sigma )\hat{\bf y},\quad 
{\bf R}'_E = - u\hat{\bf x}t  +  \frac{u}{\omega}(\ell + \sigma) \hat{\bf y}.
\end{align}
Here, we omitted the $z'$-components of the boost momentum and energy centroid, because they are ill-defined in an unbounded beam and not important for our consideration. 
Equations~\ref{EV_Bessel_moving} exhibit the transverse AM-dependent shift of the energy centroid, in agreement with Eq.~\eqref{energy_centroid} and the relativistic Hall effect \cite{Bliokh2012PRL}. One can also see that the orbital AM $\langle {\bf L}' \rangle$ acquires $\sigma$-dependent terms; this signals ``relativistic spin-orbit interaction'', i.e., the frame dependence of the spin-orbital decomposition.  

According to Eqs.~\eqref{AM_ext} and \eqref{int-ext-spin-orbit}, the extrinsic AM in the boosted frame is
\begin{equation}
\label{AM_ext_Bessel}
{\bf J}'_{\rm ext} = {\bf L}'_{\rm ext} 
\simeq u(\ell+\sigma)\left(\gamma u \hat{\bf z} + \frac{k}{\omega} \hat{\bf x} \right) .
\end{equation} 
Note that in Eq.~\eqref{AM_ext_Bessel}, as well as in expressions for $\langle {\bf J}' \rangle$ and $\langle {\bf L}' \rangle$ in Eqs.~\eqref{EV_Bessel_moving}, we omit the common time-dependent term $uk \hat{\bf y}t$, which is caused by the $x$-motion of the beam with a $z$-component of the momentum and is irrelevant to our consideration.  
Subtracting Eq.~\eqref{AM_ext_Bessel} from the total and orbital AM in Eqs.~\eqref{EV_Bessel_moving} yields their intrinsic components:
\begin{align}
\label{AM_int_Bessel}
{\bf J}'_{\rm int} & \simeq 
(\ell+\sigma)\!\left( \frac{1}{\gamma} \hat{\bf z} - u \frac{k}{\omega}\hat{\bf x}\right) = {\bf L}'_{\rm int} + \langle {\bf S}' \rangle\,, \nonumber \\
{\bf L}'_{\rm int} & \simeq  \ell\left( \frac{1}{\gamma} \hat{\bf z} - u \frac{k}{\omega}\hat{\bf x}\right) .
\end{align} 
These expressions have an elegant form with transparent physical meaning. 
In the boosted frame, ${\bf J}'_{\rm int} \parallel {\bf L}'_{\rm int} \parallel\langle {\bf S}' \rangle$, just as in the original frame. However, these intrinsic AM components are generally not parallel to the mean momentum $\langle {\bf p}' \rangle$. Only in the massless case, $m=0$, $\omega = k$, does the intrinsic AM become aligned with the momentum, similar to the spin of a massless particle, Eq.~\ref{spin_massless}:
\begin{align}
\label{AM_int_Bessel_2}
{\bf J}'_{\rm int} \simeq 
(\ell+\sigma)\frac{\langle {\bf p}'\rangle}{\langle p' \rangle}\,. 
\end{align} 

The EPL vector in the boosted frame follows from the Lorentz transformation of the EPL vector \eqref{EPL_Bessel}, ${\cal W}^\mu \simeq (\ell+s)\left(k , \omega \hat{\bf z} \right)$:
\begin{equation}
\label{EPL_Bessel_moving}
{\cal W}'^\mu \simeq 
(\ell+\sigma)\!\left(\gamma k, \omega \hat{\bf z} - \gamma u k \hat{\bf x} \right). 
\end{equation}
This agrees with the general relation ${\bf J}_{\rm int} = \bm{\mathcal W}/ \langle E \rangle$, whereas the norm of the EPL vector is given by the paraxial approximation of Eq.~\eqref{EW2_Bessel}:
\begin{equation}
\label{EW2_Beesel_paraxial}
{\cal W}^\mu{\cal W}_\mu \simeq - m^2 (\ell + \sigma)^2 \,.
\end{equation}
For $m=0$, this norm vanishes, which explains the massless-particle-like form of the intrinsic AM, Eq.~\eqref{AM_int_Bessel_2}. 

Thus, in the paraxial massless limit, the intrinsic AM of the boosted Bessel beam remains {\it purely longitudinal}, i.e., aligned with the mean momentum (despite the spatiotemporal nature of the field). 
This is in contrast to the transverse intrinsic AM found for similar Lorentz-boosted vortex beams in \cite{Bliokh2012PRA_II, Bliokh2015PR}. 
The origin of this discrepancy is the choice of the probability and energy centroids as reference points in the two approaches. Choosing the energy centroid ensures consistency between the intrinsic AM and the EPL vector.

\subsection{Spatiotemporal vortex wavepacket}
\label{sec:KG}

Finally, we combine a Bessel beam in its rest frame with a transverse Lorentz boost. This generates 2D spatiotemporal wavepackets, which have recently attracted considerable attention in optics \cite{Hancock2019O, Chong2020NP, Bliokh2021PRL, Hancock2021PRL, Bliokh2023PRA, Porras2023PIER, Wang2021O, Zang2022NP, Martin2025NP, Wan2023eL, Bekshaev2024APL} and in other wave systems \cite{Ge2023PRL, Zhang2023NC, Che2024PRL}. Here we focus exclusively on orbital AM, disregarding spin. Accordingly, the problem can be treated using scalar Klein-Gordon waves, which allow straightforward evaluation of all relevant expectation values for both massive and massless cases \cite{Bliokh2012PRL, Bliokh2012PRA_II, Bliokh2023PRA}. 

\begin{figure}[t!]
\centering
\includegraphics[width=\linewidth]{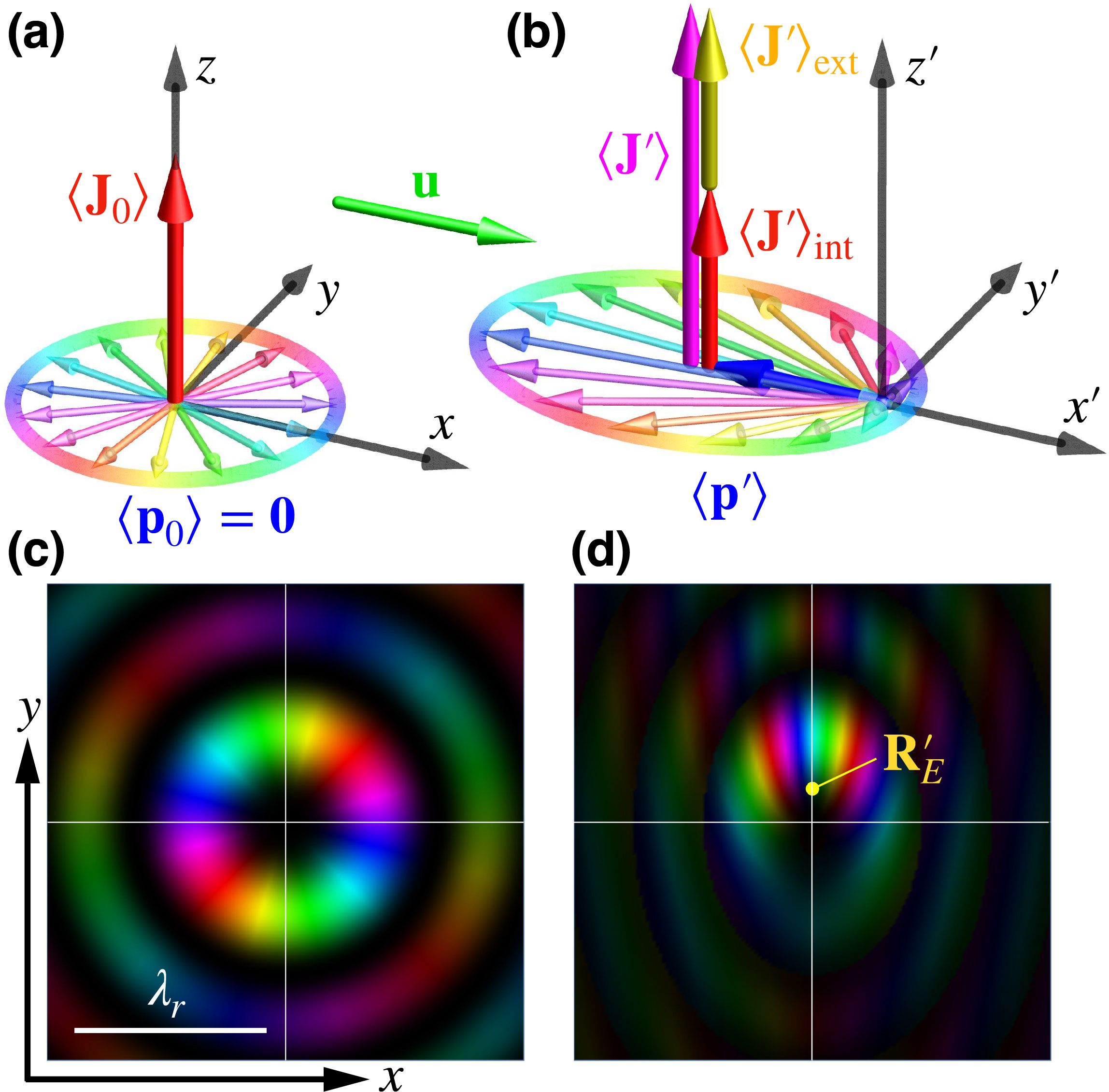}
\caption{(a) A 2D Bessel beam in its rest frame, with $k_r = k = 1.33 m$ and $\ell=2$. 
(b) Lorentz boost with velocity ${\bf u} = 0.7 \hat{\bf x}$. 
The total, intrinsic, and extrinsic AM, described by Eqs.~\eqref{EV_KG_moving} and \eqref{AM_ext_KG}, are shown by the magenta, red, and yellow arrows, respectively. The intrinsic AM is orthogonal to the mean momentum, independently of the mass $m$.
(c,d) Transverse spatial distributions of the corresponding wave fields in the rest and boosted frames. Brightness and hue colors represent the energy density and phase, respectively. The transverse shift of the energy centroid, responsible for the extrinsic AM, is indicated in (d).}
\label{Fig_4}
\end{figure}

We start with a Bessel vortex mode in its rest frame, see Fig.~\ref{Fig_4}(a). The 2D plane-wave spectrum is ${\bf k} \equiv (k_x, k_y) = k(\cos\phi, \sin\phi)$, with frequency $\omega =\sqrt{m^2 + k^2}$. Assuming a vortex phase with azimuthal index $\ell$, the expectation values of the energy, momentum, and AM are:
\begin{equation}
\label{EV_KG_rest}
\langle E_0 \rangle = \omega\,, \quad  \langle {\bf p}_0 \rangle = {\bf 0}\,, \quad
\langle {\bf J}_0 \rangle = \ell\,\hat{\bf z}\,.  
\end{equation} 
We then perform a Lorentz boost with velocity ${\bf u} = u \hat{\bf x}$, see Fig.~\ref{Fig_4}(b). 
The expectation values \eqref{EV_boost} and the energy centroid \eqref{energy_centroid} in the boosted frame become: 
\begin{align}
\label{EV_KG_moving}
\langle E \rangle & = \gamma \omega\,, \quad  \langle {\bf p} \rangle = -\gamma u \omega \hat{\bf x}\,, \quad
\langle {\bf J} \rangle = \gamma \ell\,\hat{\bf z}\,\nonumber \\
\langle {\bf K} \rangle &= -\gamma u \ell \hat{\bf y}\,, \quad
{\bf R}_E = -u \hat{\bf x}t + \frac{u\ell}{\omega}\hat{\bf y}\,.
\end{align} 
The extrinsic and intrinsic parts of the AM are obtained from Eqs.~\eqref{AM_ext} and \eqref{AM_ext}:
\begin{equation}
\label{AM_ext_KG}
{\bf J}_{\rm ext} = \gamma u^2 \ell\,\hat{\bf z}\,, \quad
{\bf J}_{\rm int} = \frac{1}{\gamma}\ell\,\hat{\bf z}\,.
\end{equation} 
This demonstrates that intrinsic orbital AM can be orthogonal or tilted with respect to the mean momentum, even for massless waves. 

The same result \eqref{AM_ext_KG} follows directly from the relation ${\bf J}_{\rm int} = \bm{\mathcal W}/ \langle E \rangle$ using the EPL vector
\begin{equation}
\label{EPL_KG}
{\cal W}^\mu = {\cal W}_0^\mu = \omega \ell (0, \hat{\bf z})\,.
\end{equation} 
Its norm is ${\cal W}^\mu {\cal W}_\mu = \omega^2 \ell^2$, which corresponds to an effectively massive case even when $m=0$.

Remarkably, the intrinsic AM in Eq.~\eqref{AM_ext_KG} is {\it twice as large} as that calculated in \cite{Hancock2021PRL, Bliokh2023PRA, Porras2023PIER} for paraxial optical (massless) spatiotemporal vortices. Indeed, the wavevector spectrum is circular in the rest frame, and it becomes elliptical after the Lorentz boost, with the ratio of the $x$ and $y$ semiaxes equal to $\gamma$, see Fig.~\ref{Fig_4}. 
Correspondingly, the real-space wave distribution becomes elliptical with semiaxes ratio of $1/\gamma$, which is a manifestation of the Lorentz contraction. For paraxial vortices with such ellipticity, the intrinsic AM (defined with respect to the energy centroid) was found to be $J_{{\rm int}\,z} \simeq \ell/2\gamma$ \cite{Hancock2021PRL, Bliokh2023PRA, Porras2023PIER}. 
This discrepancy is resolved by noticing that the wavepacket under consideration cannot be considered in the paraxial approximation. 
Although its spectrum is strongly extended along the $x$-axis for $\gamma\gg 1$, it always includes wavevectors that form large angles with the $x$ axis [see Fig.~\ref{Fig_4}(b)]; this makes the paraxial approximation inapplicable. 

This point is further clarified by the expression derived in \cite{Bliokh2023PRA} [see Eq. (21) therein]: 
\begin{equation}
\label{AM_int_KG}
{\bf J}_{\rm int} = \frac{\ell}{2} \left[ \frac{1}{\gamma} + \gamma\left( 1 - \frac{\langle p\rangle^2}{\langle E\rangle^2}\right)\right]\!\hat{\bf z}\,.
\end{equation} 
For paraxial massless waves, we assume $\langle p\rangle \simeq \langle E\rangle$, which yields the value of $\ell/2\gamma$. In our case, however, the effective mass plays a crucial role and yields $1- \langle p\rangle^2/ \langle E\rangle^2 = 1/\gamma^2$, which results in Eq.~\eqref{AM_ext_KG}.

\section{Concluding remarks}
\label{sec:conclusions}

To summarize, we have developed a framework describing the extrinsic and intrinsic AM (both spin and orbital) of relativistic wavepackets. Our approach is based on the use of the energy centroid and the expectation Pauli-Lubanski (EPL) vector, constructed from the expectation values of momentum and AM. 
In this way, it naturally links the AM treatments in classical distributed systems and in relativistic quantum mechanics. 

A notable advantage of this formalism is the absence of the ``zero-mass singularity'', characteristic of the usual PL vector. 
Indeed, any spatially-confined wavepacket necessarily has a timelike four-momentum $\langle p^\mu \rangle \langle p_\mu \rangle > 0$ and a positive effective mass. As a result, it is always possesses a rest frame, and its intrinsic AM can be arbitrary oriented with respect to the mean momentum. 

Note that the extrinsic AM, intrinsic AM, and EPL vector cannot be represented as expectation values of quantum operators; rather, they are constructed from products of expectation values. Therefore, these are properties of distributed states, which are not included in the operator structure of relativistic quantum mechanics. Nonetheless, the extrinsic and intrinsic AM, defined with respect to the energy centroid, are {separately conserved} \cite{Hancock2024PRX, Porras2024JO, Bliokh2025PLA}. This follows directly from relativistic equations of motion $d {\bf R}_E/dt = \langle {\bf p} \rangle/\langle E \rangle$, $d \langle {\bf p} \rangle/dt = {\bf 0}$, and confirms their status as physically meaningful quantities. 

In this work, we have focused primarily on the intrinsic AM ${\bf J}_{\rm int}$ calculated in a given reference frame. It can be obtained by dividing the spatial part of the EPL vector by the wavepacket energy in that frame. 
This intrinsic AM is not embedded into a four-vector and does not transform covariantly under Lorentz boosts. Alternatively, one can use covariant intrinsic AM ${\bf j}_{\rm int}$, Eq.~\eqref{AM_int_cov}, defined as the spatial part of the EPL four-vector divided by the rest-frame energy of the wavepacket. Such intrinsic AM was recently employed in \cite{Karlovets2026PRL}, and it differs from ${\bf J}_{\rm int}$ by the $\gamma$-factor relative to the rest frame.
Both quantities can be used to characterize the intrinsic AM, but it is important to keep in mind the distinction between them. 

We finally remark on the choice of the wavepacket centroid used to separate  extrinsic and intrinsic AM. In non-relativistic physics, this choice is unique, but for relativistic distributed systems the center of energy generally does not coincide with the center of probability density or center of charge \cite{Bliokh2012PRL}. In several of our earlier works, we adopted the probability centroid to define intrinsic AM \cite{Bliokh2012PRL, Smirnova2018PRA, Bliokh2023PRA}, whereas other authors have advocated the use of the energy centroid \cite{Hancock2021PRL, Hancock2024PRX, Porras2023PIER, Porras2024JO}. Although the total AM is independent of this choice (which is a matter of theoretical convenience), our present analysis demonstrates that the energy centroid provides the most consistent and elegant formulation in the context of relativistic free-space physics.

Ultimately, the relevance of a particular choice of centroid, and the corresponding definition of intrinsic AM, depends on the physical problem at hand. For example, in the interaction of an electron wavepacket with an external electromagnetic field, it is natural to expect that the {\it charge} centroid (which coincides with the probability, rather than energy, centroid) plays the primary role. Furthermore, the magnetic moment, rather than intrinsic AM, should be another relevant quantity.
By contrast, the dynamics of a relativistic wavepacket in a gravitational field is more likely to involve the energy centroid and the intrinsic AM defined in the present work. Describing such dynamics remains an important open problem for future investigation.

\begin{acknowledgments}
I am grateful to Prof. Igor P. Ivanov and Prof. Dmitry Karlovets for helpful discussions.
This work has been co-funded by the European Union through the project HORIZON-MSCA-2022-COFUND-01-SmartBRAIN3-101126600.
\end{acknowledgments}

\bibliography{refs}

\end{document}